# WAKEFIELD EXCITATION BY A SHORT LASER PULSE IN ION DIELECTRICS

*V.A. Balakirev, I.N. Onishchenko*
*NSC "Kharkov Institute of Physics and Technology", Academic Str. 1, Ukraine, 61108*
*E-mail: onish@kipt.kharkov.ua*

The process of excitation of Cherenkov electromagnetic radiation by a laser pulse in ion dielectrics is investigated. Nonlinear electric polarization in isotropic ion dielectric medium and, accordingly, polarization charges and currents induced by a ponderomotive force of a laser pulse are determined. Frequency spectra of the excited wakefields in the infrared and microwave frequency ranges are obtained. The spatiotemporal structure of the wakefield in ion dielectric waveguide is obtained and studied. It is shown that the excited field consists of a potential polarization electric field, as well as a set of eigen electromagnetic waves of ion dielectric waveguide.

## INTRODUCTION

An electric charge moving in a dielectric medium with superluminal speed radiates electromagnetic waves called Cherenkov radiation [1-4]. The electric field of a moving charge polarizes the atoms (ions) of the dielectric medium, which in turn coherently re-radiate electromagnetic waves.

A similar effect takes place when a high-power laser pulse propagates in a dielectric [5-10]. A necessary condition for the appearance of a Cherenkov radiation of a laser pulse is that the group velocity of the laser pulse must exceed the phase velocity of the radiated electromagnetic wave. The effect of Cherenkov radiation of a laser pulse in a dielectric medium is as follows. When a laser pulse propagates in a dielectric a pulsed ponderomotive force quadratic in the laser field propagating in the medium with the group velocity of the laser pulse will act on the bonded electrons of the atoms (ions) of a medium. This force, in turn, will lead to the polarization of the atoms (ions) of the dielectric. Induced polarization charges and currents will coherently radiate electromagnetic waves (Cherenkov radiation). The effect of the Cherenkov radiation of a laser pulse is quite similar to the Cherenkov radiation of an electron bunch moving in a dielectric medium, with the difference that the ponderomotive force of the laser pulse plays the role of the pulse electric field of the electron bunch. Note that when the laser pulse crosses the dielectric boundary, the transition radiation effect is possible [7] by analogy with the case of an electron bunch [11].

The Cherenkov wakefield radiation in a dielectric medium caused by a high-power ultrashort laser pulse can be used to accelerate charged particles similarly to a laser-plasma wakefield acceleration method [12-14].

In [7, 9], the effect of the Cherenkov radiation of a laser pulse was studied using a simple model of a dielectric medium consisting of atoms of the same type. A bright example of such a medium is diamond, whose crystal lattice consists only of carbon atoms. The carbon atoms in diamond are held by covalent forces, which are of a quantum nature and arise as a result of the bonding pairs of the valence electrons of neighboring atoms (overlapping of the wave functions of the valence electrons). Atoms retain their electrical neutrality. Only the electron shells of atoms contribute to the electric polarization of covalent dielectrics. Due to large mass the nuclei of atoms do not participate in the polarization of dielectrics. Namely, due to the electronic nature of polarization, for covalent dielectrics, the values of dielectric constant in the optical frequency range and in the static limit are close.

A much wider class of dielectrics is formed by ion-bonded dielectrics [15-19]. No pure element of the periodic table is related to dielectrics of this class. All ion dielectrics are chemical compounds. Ion crystals are composed of positive and negative ions. These ions form a crystal lattice as a result of Coulomb attraction of oppositely charged ions.

The traditional examples of ion dielectrics are crystals of an alkali-halide group with the formula $A_I B_{VII}$ (for example, NaCl and KCl). In crystals of this group, it is energetically advantageous for an atom of alkali metal to transfer its valence electron to an adjacent halide atom and fill its outer shell. As a result, an ion bond arises between the atoms of different elements. This bond is due to the interaction of oppositely charged ions. Below we restrict consideration to the simplest case of diatomic crystals. These dielectrics also include ion crystals with the formulas $A_{II} B_{VI}$ and $A_{III} B_V$. Note also that in ion crystals a covalent bond share is always present. For example, in ion crystals of the alkali halide group, in the total binding energy, it is less than 6% [19].

In determining the total electric polarization induced by a laser pulse in an ion dielectric, it is necessary to take into account both the total contribution of the polarizations of the electron shells of all the ions which form the crystal and the total contribution of the positive and negative ions of the crystal.

In this paper, a system of nonlinear equations of macroscopic electrodynamics is formulated, which describes the process of excitation of Cherenkov radiation by a laser pulse in an ion dielectric medium. On the basis of these equations, the effect of the Cherenkov radiation of a laser pulse in a dielectric waveguide (light guide) will be investigated. A complete picture of the excitation of Cherenkov radiation by a laser pulse propagating in an ion dielectric is presented. The frequency spectrum of Cherenkov radiation is determined. The spatio-temporal structure of the Cherenkov electromagnetic field has been obtained and studied.

## 1. PROBLEM STATEMENT. BASIC EQUATIONS

We consider the laser pulse (wave packet) propagation in a homogeneous dielectric medium. Laser pulse electromagnetic field components are

$$\vec{E}_L(\vec{r},t) = \frac{1}{2}\vec{E}_{L0}(\vec{r},t)e^{i\psi_L} + c.c.,$$

$$\vec{H}_L(\vec{r},t) = \frac{1}{2ik_0} rot\left[\vec{E}_{L0}(\vec{r},t)e^{i\psi_L}\right] + c.c., \quad (1)$$

where $\psi_L = \vec{k}_L\vec{r} - \omega_L t$, $\vec{k}_L$ is wave vector, $k_0 = \omega_L/c$, $\omega_L$ is carrier frequency of a laser pulse, $\vec{E}_{L0}(\vec{r},t)$ is a laser pulse envelope slowly varying in space and time.

Under the action of the ponderomotive force (high frequency pressure force) of the laser pulse it is arisen a slow on the carrier frequency scale polarization in the dielectric, which in turn becomes the source of the electromagnetic field (Cherenkov radiation). Maxwell's system of equations describing the electromagnetic field, which is excited by a polarization induced by a laser pulse, has the form

$$rot\vec{E} = -\frac{1}{c}\frac{\partial \vec{H}}{\partial t}, \quad rot\vec{H} = \frac{1}{c}\frac{\partial \vec{E}}{\partial t} + \frac{4\pi}{c}\frac{\partial \vec{P}}{\partial t},$$

$$div\vec{E} = -4\pi div\vec{P}, \quad div\vec{H} = 0, \quad (2)$$

$\vec{P}$ is vector of electric polarization.

In ion dielectrics there are two mechanisms of electric polarization. This is primarily an electronic polarization mechanism inherent in all types of dielectrics. Electron polarization is due to the displacement of a shell of bound electrons relative to their nuclei under the action of an electric field. The second polarization mechanism is ionic; as it is caused by the relative displacement of oppositely charged ions. It should be noted that such a separation of the polarization mechanisms is not quite rigorous. A more adequate is the polarization model, in which the ions are not only displaced, but also deformed (the model of deformable ions [16]). Under the action of an electric field, the electron shell of each ion will be deformed and displaced relative to the nucleus, so that an internal dipole moment is formed in the ion, which will weaken the applied electric field. Accordingly, the force causing the displacement of the ions will be decreased and, as a result, the ion polarization will be decreased. Qualitatively, this weakening effect can be taken into account by renormalizing the ion charge or by introducing the effective Scigetty charge [17]. For most ion dielectrics, the Scigetti charge is 0.7–0.9 of the ion charge. However, to simplify the analysis of the Cherenkov effect of a laser pulse, we restrict ourselves to the model of hard (non-deformable) ions.

First of all, we formulate equations describing the electron polarization of diatomic ion crystals induced by a laser pulse. Induced electron polarization can be described in the framework of the following model. An atom is represented as a point nucleus surrounded by an electron cloud. When the electron cloud is displaced as a whole relative to the nucleus, a dipole moment of the atom $\vec{p} = -Ze\vec{r}$ arises, where $\vec{r}$ is the radius-vector of the electron cloud center, $Ze$ is nucleus charge. Accordingly, a dipole returning force will act on the cloud [17]

$$\vec{F}_{de} = -\frac{(Ze)^2}{R_0^3}\vec{r},$$

which leads to harmonic dipole oscillations of an atom with its eigen frequency

$$\omega_{de} = \sqrt{\frac{Ze^2}{mR_0^3}}, \quad (3)$$

$R_0$ is the radius of the atom. Relation (3) allows us to estimate the frequency of dipole oscillations of individual ions. For most elements, this frequency lies in the optical range.

In a condensed medium, each atom is in a local (acting) electric field $\vec{E}_{loc}$, which can differ substantially from the macroscopic field $\vec{E}$ included in Maxwell's equations (2). The local electric field $\vec{E}_{loc}$ includes both the external field and the total electric field of the induced dipoles surrounding a given atom (ion). In a crystal medium with a cubic crystal lattice, the local electric field is described by the Lorentz formula [15,16]

$$\vec{E}_{loc} = \vec{E} + \frac{4\pi}{3}\vec{P}. \quad (4)$$

Based on the model of a "hard" electron shell, we present the motion equation of the electron shell relatively to the nuclei of the ions of crystal in the form

$$\frac{d^2\vec{R}_{e\alpha}}{dt^2} + \omega_{de\alpha}^2 \vec{R}_{e\alpha} = -\frac{e}{m}\left(\vec{E}_{tot}^{loc}(\vec{R}_{e\alpha},t) + \frac{1}{c}\left[v_{e\alpha}\vec{H}_{tot}\right]\right), \quad (5)$$

where $\vec{R}_{e\alpha}$ is radius-vector of the mass center of the electron shell of the ion, $\alpha = \pm$, signs $+,-$ correspond to positive and negative ions, $v_{e\alpha}(t) = d\vec{R}_{e\alpha}/dt$ are velocities of electron shells of ions, $\omega_{de\alpha}$ are frequencies of dipole oscillations of the electron shells of positive and negative ions. Total local electric field

$$\vec{E}_{tot}^{loc} = \vec{E}_L^{loc} + \vec{E}_{loc}$$

includes both the local electric field of the laser pulse

$$\vec{E}_L^{loc} = \vec{E}_L + \frac{4\pi}{3}\vec{P}_L,$$

and the local electric field (4), containing a slow (Cherenkov) field $\vec{E}$ included in Maxwell's equations (2), $\vec{P}_L$ is the electron polarization at the carrier frequency of the laser pulse. Total magnetic field is

$$\vec{H}_{tot} = \vec{H}_L + \vec{H}.$$

The polarization $\vec{P}_L$ is related to the laser electric field by the relation $\vec{P}_L = \chi_L \vec{E}_L$, where $\chi_L = (\varepsilon_L - 1)/4\pi$ is the electric susceptibility and the $\varepsilon_L$ is dielectric constant of the ion dielectric at the laser frequency. Given these relations for the local laser electric field, we obtain the following expression [18]

$$\vec{E}_L^{loc} = \frac{\varepsilon_L + 2}{3} \vec{E}_L.$$

We will seek an approximate solution of the motion equation (5) in the form of the sum

$$\vec{R}_{e\alpha}(t) = \vec{R}_{eL}^{(\alpha)}(t) + \vec{R}_{e0}^{(\alpha)}(t),$$

quickly oscillating at the laser frequency displacement and slow displacement $\vec{R}_{eL}^{(\alpha)}(t)$ of the electron cloud relatively to the ion nucleus. A quickly oscillating displacement is described by the linear equation of motion of the oscillator

$$\frac{d^2 \vec{R}_{eL}^{(\alpha)}}{dt^2} + \omega_{de\alpha}^2 \vec{R}_{eL}^{(\alpha)} = \\ = -\frac{e}{2m} \frac{\varepsilon_L + 2}{3} \vec{E}_{L0}(\vec{R}_{e0}^{(\alpha)}, t) e^{-i\left(\omega_L t - \vec{k}_L \vec{R}_{e0}^{(\alpha)}\right)} + c.c.. \quad (6)$$

From this equation of motion we find expressions for the quickly oscillating displacement and velocity of the electron shell of each ion

$$\vec{R}_{eL}^{(\alpha)} = -a^{(\alpha)} \frac{1}{2} \left[ \vec{E}_{L0}(\vec{R}_{e0}^{(\alpha)}, t) e^{-i\left(\omega_L t - \vec{k}_L \vec{R}_{e0}^{(\alpha)}\right)} + c.c. \right], \quad (7a)$$

$$\vec{v}_{eL}^{(\alpha)} = \omega_L a^{(\alpha)} \frac{1}{2} \left[ i\vec{E}_{L0}(\vec{R}_{e0}^{(\alpha)}, t) e^{-i\left(\omega_L t - \vec{k}_L \vec{R}_{e0}^{(\alpha)}\right)} + c.c. \right], \quad (7b)$$

$$a^{(\alpha)} = \frac{e}{m} \frac{\varepsilon_L + 2}{3} \frac{1}{\omega_{de\alpha}^2 - \omega_L^2}, \quad \alpha = +, -.$$

We now formulate the equation for the slow displacements of electronic dipole oscillators. We use the expansion of the laser electric field

$$\vec{E}_{L0}(\vec{R}_{e0}^{(\alpha)} + \vec{R}_{eL}^{(\alpha)}, t) = \vec{E}_{L0}(\vec{R}_{e0}^{(\alpha)}, t) + \left(\vec{R}_{eL}^{(\alpha)} \nabla\right) \vec{E}_{L0}(\vec{R}_{e0}^{(\alpha)})$$

and we will save in the right side of the motion equation (5) only terms quadratic on the laser field. As a result, we obtain the following equation for the slow displacement of the electron shell of an individual ion

$$\frac{d^2 \vec{R}_{e0}^{(\alpha)}}{dt^2} + \omega_{de\alpha}^2 \vec{R}_{e0}^{(\alpha)} = -\frac{e}{m} \left( \vec{E} + \frac{4\pi}{3} \vec{P} + \\ + \frac{\varepsilon_L + 2}{3} \left\langle \left(\vec{R}_{eL}^{(\alpha)} \nabla\right) \vec{E}_L(\vec{R}_{e0}^{(\alpha)}) \right\rangle + \frac{1}{c} \left\langle \left[ \vec{v}_{eL}^{(\alpha)} \vec{H}_L \right] \right\rangle \right). \quad (8)$$

Angle brackets mean averaging over fast oscillations at the frequency of the laser pulse. Performing the averaging procedure with taking into account expressions (1), (7a), (7b) for the quantities included in (8), we obtain the equation for the slow displacement of the electron shell

$$\frac{d^2 \vec{R}_{e0}^{(\alpha)}}{dt^2} + \omega_{de\alpha}^2 \vec{R}_{e0}^{(\alpha)} = -\frac{e}{m} \left( \vec{E} + \frac{4\pi}{3} \vec{P} \right) + \frac{1}{m} \vec{F}_{pon}^{(\alpha)}, \quad (9)$$

$$\vec{F}_{pon}^{(\alpha)} = \frac{e^2}{4m} \frac{\varepsilon_L + 2}{3} \frac{1}{\omega_{de\alpha}^2 - \omega_L^2} \vec{\Pi}, \quad (10)$$

$$\vec{\Pi} = \nabla \left| \vec{E}_{L0} \right| + \frac{\varepsilon_L - 1}{3} \left[ \left(\vec{E}_{L0} \nabla\right) \vec{E}_{L0}^* + \left(\vec{E}_{L0}^* \nabla\right) \vec{E}_{L0} \right], (11)$$

$\vec{F}_{pon}^{(\alpha)}$ is ponderomotive force acting on the electron shell of each ion.

The first term in (11) describes the gradient force of the high frequency pressure. The second term arises only for condensed media and is due to the difference between a local laser electric field in crystal acting on a single ion and a macroscopic field in a medium that obeys Maxwell's equations. In dielectric media where the acting field coincides with the external field, for example, in the gas dielectric or plasma this term is absent.

Under the action of ponderomotive force in dielectric electron polarization appears

$$\vec{P}_e = \vec{P}_e^{(+)} + \vec{P}_e^{(-)},$$

where $\vec{P}_e^{(\alpha)} = -q^{(\alpha)} N_0 \vec{R}_{e0}^{(\alpha)}$ are the partial electron polarizations of positive ($\alpha = +$) and negative ($\alpha = -$) ions, $q^{(\alpha)}$ are full charges of electron shells of corresponding ions, $N_0$ is the concentration of ions of each type.

Partial electron polarizations are described by the following equations, which follow directly from the equations of motion (9)

$$\frac{\partial^2 \vec{P}_e^{(\pm)}}{\partial t^2} + \omega_{de(\pm)}^2 \vec{P}_e^{(\pm)} - \frac{1}{3} \omega_{pe(\pm)}^2 \vec{P} = \frac{1}{4\pi} \omega_{pe(\pm)}^2 \vec{E} - \\ - \frac{eN_0}{4m} \frac{\varepsilon_L + 2}{3} \alpha_L^{(\pm)} \vec{\Pi}, \quad (12)$$

where

$$\alpha_L^{(\pm)} = \frac{eq^{(\pm)}}{m} \frac{1}{\omega_{de(\pm)}^2 - \omega_L^2}$$

are electron polarizabilities of individual positive and negative ions at laser pulse frequencies, $\omega_{pe(\pm)}^2 = \frac{4\pi e q^{(\pm)} N_0}{m}$ is square of the effective plasma frequency.

The left-hand sides of equations (12) for electron polarizations include complete polarization of the ion dielectric.

$$\vec{P} = \vec{P}_e^{(+)} + \vec{P}_e^{(-)} + \vec{P}_i, \quad (13)$$

which also includes ion polarization $\vec{P}_i$. Ion polarization occurs as a result of the relative displacement of positive and negative ions under the action of an electric field. If ions are not deformed, then the dipole moment of the unit cell of the crystal containing two ions of opposite sign is

$$\vec{p}_i = q_i \vec{R}_i, \quad \vec{R}_i = \vec{R}_i^{(+)} - \vec{R}_i^{(-)},$$

where $\vec{R}_i^{(\pm)}$ are the displacements of positive and negative ions from the equilibrium position, $q_i$ is ion charge. If the crystal deformation is smooth over the microscopic scale of the crystal (unit cell size), then the displacements of positive and negative ions obey to the equations

$$M^{(+)} \frac{d^2 \vec{R}_i^{(+)}}{dt^2} + K \left( \vec{R}_i^{(+)} - \vec{R}_i^{(-)} \right) = q_i \left( \vec{E} + \frac{4\pi}{3} \vec{P} \right),$$

$$M^{(-)} \frac{d^2 \vec{R}_i^{(-)}}{dt^2} + K \left( \vec{R}_i^{(-)} - \vec{R}_i^{(+)} \right) = -q_i \left( \vec{E} + \frac{4\pi}{3} \vec{P} \right), (14)$$

which are reduced to one equation for the relative displacement of ions

$$\frac{d^2 \vec{R}_i}{dt^2} + \omega_{di}^2 \vec{R}_i = \frac{q_i}{M} \left( \vec{E} + \frac{4\pi}{3} \vec{P} \right), \quad (15)$$

$M^{(\pm)}$ are ion masses, $K$ is force parameter, $\omega_{di} = \sqrt{K/M}$ is the eigen frequency of ion dipole oscillations, $M = M^{(+)} M^{(-)} / (M^{(+)} + M^{(-)})$ is reduced mass. Note that since the ponderomotive force acting on ions is inversely proportional to the mass of the ion, then it is small and we neglected it in equation (14),(15). Equation of motion (15) implies the following equation for ion polarization

$$\frac{\partial^2 \vec{P}_i}{\partial t^2} + \omega_{di}^2 \vec{P}_i - \frac{1}{3}\omega_{pi}^2 \vec{P} = \frac{1}{4\pi}\omega_{pi}^2 \vec{E},$$

where $\omega_{pi}^2 = \frac{4\pi q_i^2 N_0}{M}$ is the square of the ion plasma frequency.

Thus, partial polarizations are described by a system of equations of coupled linear oscillators [9].

$$\frac{\partial^2 \vec{P}_e^{(+)}}{\partial t^2} + \omega_{de+}^2 \vec{P}_e^{(+)} - \frac{1}{3}\omega_{pe+}^2 \left(\vec{P}_e^{(+)} + \vec{P}_e^{(-)} + \vec{P}_i\right) =$$
$$= \frac{1}{4\pi}\omega_{pe+}^2 \vec{E} - \frac{eN_0}{4m}\frac{\varepsilon_L + 2}{3}\alpha_L^{(+)}\vec{\Pi},$$

$$\frac{\partial^2 \vec{P}_e^{(-)}}{\partial t^2} + \omega_{de-}^2 \vec{P}_e^{(-)} - \frac{1}{3}\omega_{pe-}^2 \left(\vec{P}_e^{(+)} + \vec{P}_e^{(-)} + \vec{P}_i\right) =$$
$$= \frac{1}{4\pi}\omega_{pe-}^2 \vec{E} - \frac{eN_0}{4m}\frac{\varepsilon_L + 2}{3}\alpha_L^{(-)}\vec{\Pi}, \quad (16)$$

$$\frac{\partial^2 \vec{P}_i}{\partial t^2} + \omega_{di}^2 \vec{P}_i - \frac{1}{3}\omega_{pi}^2 \left(\vec{P}_e^{(+)} + \vec{P}_e^{(-)} + \vec{P}_i\right) = \frac{1}{4\pi}\omega_{pi}^2 \vec{E}.$$

The external force exciting these oscillators is the ponderomotive force from the side of the laser pulse.

The Maxwell equations (2), together with the equations for partial polarizations (16) and the relation (13) for the full polarization, are completely closed and describe the Cherenkov excitation of electromagnetic radiation of a laser pulse in an ion dielectric.

We will solve this system of equations by the method of Fourier transform

$$\vec{E}(\vec{r},t) = \int_{-\infty}^{\infty} \vec{E}_\omega(r)e^{-i\omega t}d\omega, \ \vec{P}(\vec{r},t) = \int_{-\infty}^{\infty} \vec{P}_\omega(r)e^{-i\omega t}d\omega,$$

where, $\vec{E}_\omega(r), \vec{P}_\omega(\vec{r})$ are Fourier-components of the corresponding quantities. For example

$$\vec{E}_\omega(\vec{r}) = \frac{1}{2\pi}\int_{-\infty}^{\infty} \vec{E}(r,t)e^{i\omega t}dt.$$

From the system of coupled equations for partial polarizations (16) we find the expression for the Fourier components of the full polarization vector

$$\vec{P}_\omega = \frac{\varepsilon(\omega)-1}{4\pi}\vec{E}_\omega - \mu\vec{\Pi}_\omega, \quad (17)$$

where

$$\mu = \frac{eN_0}{4m}\frac{\varepsilon_L + 2}{3}\frac{\varepsilon(\omega)+2}{3}\Gamma(\omega),$$

$$\varepsilon(\omega) = \frac{1 + \frac{2}{3}\Lambda(\omega)}{1 - \frac{1}{3}\Lambda(\omega)}, \quad (18)$$

$$\Lambda(\omega) = \frac{\omega_{pi}^2}{\omega_{di}^2 - \omega^2} + \frac{\omega_{pe-}^2}{\omega_{de-}^2 - \omega^2} + \frac{\omega_{pe+}^2}{\omega_{de+}^2 - \omega^2},$$

$$\Gamma(\omega) = \frac{\alpha_L^{(+)}}{\omega_{de+}^2 - \omega^2} + \frac{\alpha_L^{(-)}}{\omega_{de-}^2 - \omega^2}.$$

$\vec{\Pi}_\omega$ is Fourier-component of the ponderomotive force (10). The value $\varepsilon(\omega)$ is the dielectric constant of a diatomic dielectric with an ion bond. Note that the Lorentz-Lorentz formula follows from the expression for permittivity [15,16]

$$\frac{\varepsilon(\omega)-1}{\varepsilon(\omega)+2} = \frac{4\pi}{3}N_0\left(\alpha_e^{(-)} + \alpha_e^{(+)} + \alpha_i\right), \quad (19)$$

where

$$\alpha_e^{(\pm)}(\omega) = \frac{eq^{(\pm)}}{m}\frac{1}{\omega_{de(\pm)}^2 - \omega^2}$$

are electron polarizabilities of ions,

$$\alpha_i(\omega) = \frac{q_i^2}{M}\frac{1}{\omega_{di}^2 - \omega^2}$$

is ion polarizability of a pair of oppositely charged ions in the unit cell. The relation (19) establishes a relationship between the dielectric constant and the sum of the polarizations of all particles forming the crystal.

Maxwell's system of equations for Fourier-component of the electromagnetic field with taking into account the relation for the full polarization (17) can be represented as

$$rot\vec{H}_\omega = -ik_0\varepsilon(\omega)\vec{E}_\omega + \frac{4\pi}{c}\vec{j}_{\omega pol}, \ rot\vec{E}_\omega = ik_0\vec{H}_\omega,$$

$$\varepsilon(\omega)div\vec{E}_\omega = 4\pi\rho_{\omega pol}, \quad div\vec{H}_\omega = 0, \quad (20)$$

$k_0 = \omega/c$. The Fourier-components of the polarization currents and charges induced in the ion dielectric by the ponderomotive force of a laser pulse are described by the expressions

$$\vec{j}_{pol\omega} = i\omega\mu\vec{\Pi}_\omega, \quad \rho_{pol\omega} = \mu div\vec{\Pi}_\omega. \quad (21)$$

The resulting working system of equations makes it possible to investigate Cherenkov radiation in a wide variety of physical situations: the model of an infinite dielectric medium, dielectric waveguides and cavities.

## 2. CHERENKOV RADIATION OF A LASER PULSE IN A DIELECGRIC WAVEGUIDE

We consider the dielectric waveguide, made in the form of a homogeneous dielectric cylinder, the lateral surface of which is covered with a perfectly conductive metal film. A circularly polarized laser pulse with electric field components propagates along the axis of the waveguide

$$E_{0x} = \sqrt{\frac{I_0}{2}}\psi(r,\tau), \ E_{0y} = iE_{0x}, \quad (22)$$

$$\psi(r,\tau) = \left[R(r/r_L)T(\tau/t_L)\right]^{1/2}.$$

The function $R(r/r_L)$ describes the radial profile of the laser pulse intensity $I_0 = \left|\vec{E}_0\right|^2$, $R(0) = 1$, $R(r = b) = 0$, $b$ is the waveguide radius, the function $T(\tau)$ describes the longitudinal profile, $\tau = t - z/v_g$, $v_g$ is the group velocity, $maxT(\tau) = 1$, $I_0$ is the maximum intensity, $r_L$ is the radius of the laser pulse, $t_L$ is pulse duration. We will describe the laser wave beam in the approximation of a fixed cross section and do not take into account the diffraction expansion of the wave beam in the transverse direction.

From the system of Maxwell equations (20) the wave equation for the longitudinal Fourier component of the Cherenkov electric field is the following

$$\Delta E_{z\omega} + k_0^2 \varepsilon(\omega) E_{z\omega} = 4\pi \left( \frac{1}{\varepsilon(\omega)} \frac{\partial \rho_{pol\omega}}{\partial z} - i\frac{k_0}{c} j_{zpol\omega} \right). \quad (23)$$

Fourier-components of polarization charges and currents $\rho_{pol\omega}$, $j_{zpol\omega}$ are defined by expressions (21). For a circularly polarized laser pulse (22), these expressions take the form

$$\rho_{pol\omega} = \mu \left[ (\Delta_\perp - k_g^2) I_\omega(r) + \frac{\varepsilon_L - 1}{6} \Delta_\perp I_\omega(r) \right] e^{ik_g z}, \quad (24)$$

$$j_{zpol\omega} = -\omega k_g \mu I_\omega(r) e^{ik_g z}, \quad (25)$$

where $k_g = \omega/v_g$, $\Delta_\perp$ is the transverse part of Laplacian, $I_\omega(r)$ is Fourier component of the intensity of the laser pulse field. We introduce a function

$$D_{\omega z} = \varepsilon(\omega) E_z - 4\pi i k_g \mu I_\omega(r). \quad (26)$$

For this function, instead of equation (23), taking into account relations (24), (25), we obtain the equation

$$\Delta D_{\omega z} + k_0^2 \varepsilon(\omega) D_{\omega z} = 4\pi i k_g \mu \frac{\varepsilon_L - 1}{6} e^{ik_g z} \Delta_\perp I_\omega(r). \quad (27)$$

The function $D_{\omega z}$ has a simple physical meaning and is a longitudinal Fourier-component of the longitudinal electric displacement field $D_z = E_z + 4\pi P_z$, taking into account the polarization (17) caused by the action of the ponderomotive force of the laser pulse.

The longitudinal component of electric displacement field should be sought as a series of Bessel functions.

$$D_{\omega z} = e^{ik_g z} \sum_{n=0}^{\infty} C_n(\omega) J_0\left( \lambda_n \frac{r}{b} \right), \quad (28)$$

where $\lambda_n$ are the roots of the Bessel function $J_0(x)$. Using the orthogonality of the Bessel functions $J_0(\lambda_n r/b)$, from the equation (27) we find the expansion coefficients

$$C_n(\omega) = -4\pi i k_g \mu \frac{\varepsilon_L - 1}{6} I_0 T(\omega) \frac{\delta_n}{\Delta_n(\omega)}. \quad (29)$$

Here

$$\delta_n = \frac{\lambda_n^2}{b^2} \frac{\rho_n}{N_n}, N_n = \frac{b^2}{2} J_1^2(\lambda_n), \rho_n = \int_0^b R\left(\frac{r}{r_L}\right) J_0\left( \lambda_n \frac{r}{b} \right) r dr,$$

$$T(\omega) = \frac{1}{2\pi} \int_{-\infty}^{\infty} T(\tau/t_L) e^{i\omega\tau} d\tau,$$

$$\Delta_n(\omega) = k_0^2 \varepsilon(\omega) - \frac{\omega^2}{v_g^2} - \frac{\lambda_n^2}{b^2}. \quad (30)$$

Taking into account relations (26), (28), we obtain the following expression of Fourier component of the longitudinal electric field

$$E_{\omega z}(r) = A_0 T(\omega) G(r, \omega) e^{ik_g z}. \quad (31)$$

Here

$$A_0 = \frac{\pi e N_0 I_0}{m v_g} \frac{\varepsilon_L + 2}{3},$$

$$G(r, \omega) = i\omega \frac{\varepsilon(\omega) + 2}{3\varepsilon(\omega)} \Gamma(\omega) \Phi_\omega(r) -$$

$$\frac{i\omega(\varepsilon_L - 1)[\varepsilon(\omega) + 2]}{18} \Gamma(\omega) \sum_{n=1}^{\infty} \delta_n \frac{k_0^2}{k_n^2 \Delta_n(\omega)} J_0\left( \lambda_n \frac{r}{b} \right), \quad (32)$$

$$\Phi_\omega(r) = R(r) + \frac{\varepsilon_L - 1}{6} \sum_{n=1}^{\infty} \frac{\delta_n}{k_n^2} J_0\left( \lambda_n \frac{r}{b} \right), k_n^2 = k_g^2 + \lambda_n^2/b^2.$$

Accordingly, the longitudinal component of the excited electric field can be represented as a convolution

$$E_z(r, \tau) = A_0 \frac{1}{2\pi} \int_{-\infty}^{\infty} T(\tau_0/t_L) G(r, \tau - \tau_0) d\tau_0, \quad (33)$$

where

$$G(r, \tau - \tau_0) = \frac{1}{2\pi} \int_{-\infty}^{\infty} G(r, \omega) e^{-i\omega(\tau - \tau_0)} d\omega \quad (34)$$

is Green function. For further analysis, we will present the Green function in the form

$$G(r, \tau - \tau_0) = G_l(r, \tau - \tau_0) + G_{tr}(r, \tau - \tau_0), \quad (35)$$

$$G_l(r, \tau - \tau_0) = i \int_{-\infty}^{\infty} \frac{\varepsilon(\omega) + 2}{3} \Gamma(\omega) \Phi_\omega(r) \frac{\omega}{\varepsilon(\omega)} e^{-i\omega(\tau - \tau_0)} d\omega,$$

$$G_{tr}(r, \tau - \tau_0) = -\frac{\varepsilon_L - 1}{6} \sum_{n=1}^{\infty} \delta_n J_0\left( \lambda_n \frac{r}{b} \right) S_n(\tau - \tau_0),$$

$$S_n(\tau - \tau_0) = i \int_{-\infty}^{\infty} \frac{\varepsilon(\omega) + 2}{3} \Gamma(\omega) \frac{k_0^2}{k_n^2} \frac{\omega}{\Delta_n(\omega)} e^{-i\omega(\tau - \tau_0)} d\omega.$$

The Green function actually describes the structure of the wakefield in a dielectric medium excited by a laser pulse with $\delta$-shaped longitudinal intensity profile. Moreover, the term $G_l(r, \tau - \tau_0)$ takes into account the excitation of potential longitudinal oscillations of the ionic dielectric, and the term $G_{tr}(r, \tau - \tau_0)$ describes the excitation of transverse electromagnetic waves.

## 2.1. FREQUENCY DISPERSION OF DIELECTRIC PERMEABILITY

The Green's function (34) and, accordingly, the wakefield (33) are largely determined by the value and frequency dispersion of the dielectric constant $\varepsilon(\omega)$ determined by the formula (18). For the qualitative analysis of this dependence, the expression for the dielectric constant can be conveniently represented as

$$\varepsilon(\omega) = \frac{(\omega^2 - \omega_{Li}^2)(\omega^2 - \omega_{Le-}^2)(\omega^2 - \omega_{Le+}^2)}{(\omega^2 - \omega_{Ti}^2)(\omega^2 - \omega_{Te-}^2)(\omega^2 - \omega_{Te+}^2)}, \quad (36)$$

where $\omega_{Li}, \omega_{Le(\pm)}$ are the roots of the cubic equation with respect to the square of the frequency

$$\Lambda(\omega) = -\frac{3}{2}. \quad (37)$$

Fig.1 shows the graph of the qualitative dependence $\Lambda(\omega)$. The horizontal line $\Lambda = -3/2$ defines the roots of equation (37), i.e. frequency spectrum of potential longitudinal oscillations of an ion crystal. All three roots are positive, i.e. the frequencies are real. At these frequencies, the dielectric permeability is zero. The frequency $\omega_{Li}$ is the frequency of longitudinal optical phonons and belongs to the infrared frequency range. Frequencies $\omega_{Le(\pm)}$ are the frequencies of longitudinal polarization electron oscillations and are in the optical or even ultraviolet frequency ranges. The specified frequencies are in the intervals

$$\omega_{de-} > \omega_{Li} > \omega_{di}, \omega_{de+} > \omega_{Le-} > \omega_{de-}, \omega_{de-} < \omega_{Le+}.$$

For definiteness, we assumed that $\omega_{de+} > \omega_{de-}$. Since the frequencies of the longitudinal ion and electron oscillations are very different, the roots of the cubic equation (37) can be found approximately

$$\omega_{Li}^2 = \omega_{di}^2 + \frac{2}{9}\frac{\varepsilon_{opt}+2}{\varepsilon_{opt}}\omega_{pi}^2, \qquad (38)$$

$$\omega_{Le(\pm)}^2 = \frac{1}{2}\left[\omega_{ge+}^2 + \omega_{ge-}^2 \pm \sqrt{\left(\omega_{ge+}^2 + \omega_{ge-}^2\right)^2 + \frac{16}{9}\omega_{pe+}^2\omega_{pe-}^2}\right].$$

Here

$$\omega_{ge(\pm)}^2 = \omega_{de(\pm)}^2 + \frac{2}{3}\omega_{pe(\pm)}^2,$$

$$\varepsilon_{opt} = \frac{1+\frac{2}{3}\Lambda_{opt}}{1-\frac{1}{3}\Lambda_{opt}}, \qquad \Lambda_{opt} = \frac{\omega_{pe+}^2}{\omega_{de+}^2} + \frac{\omega_{pe-}^2}{\omega_{de+}^2},$$

$\varepsilon_{opt}$ is dielectric permeability of an ion crystal in the optical frequency range

$$\omega_{de+}^2 >> \omega^2 >> \max(\omega_{pi}^2, \omega_{di}^2).$$

The poles of the dielectric constant (36) are the roots of the cubic equation with respect to the square of the frequency

$$\Lambda(\omega) = 3. \qquad (39)$$

Fig.1 illustrates the location of these roots. The values of these roots are determined by the intersection points of the curve $\Lambda(\omega)$ and the horizontal line $\Lambda = 3$.

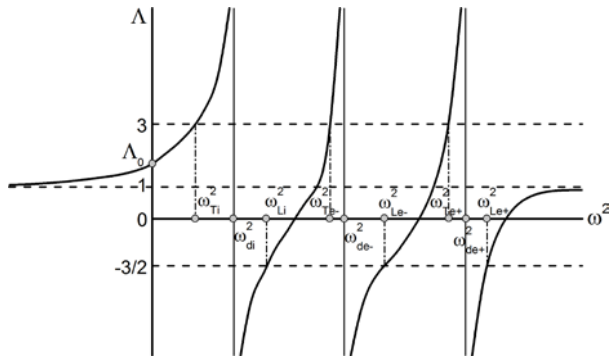

*Fig.1. Graph of function $\Lambda(\omega^2)$.*

Cubic equation has three positive roots too. These roots correspond to the frequencies $\omega_{Ti}, \omega_{Te(\pm)}$. These frequencies are the absorption lines of the electromagnetic waves of an ion crystal. In the vicinity of these frequencies, the imaginary part of the dielectric constant and, accordingly, the energy losses of electromagnetic waves increase greatly. The frequency of absorption by the ion subsystem is the frequency of transverse optical phonons. Note that the optical longitudinal and transverse ion oscillation branches are characterized by the fact that in the unit cell of the crystal oppositely charged ions are displaced towards each other. At the same time, the center of gravity of the unit cell remains motionless. As in the case of longitudinal optical phonons, the frequencies of transverse optical phonons lie in the infrared range. Electron resonance absorption frequencies are in the optical ranges. For the indicated frequencies from the cubic equation (39) we find the following approximate expressions

$$\omega_{Ti}^2 = \omega_{di}^2 \frac{3-\Lambda_{st}}{3-\Lambda_{opt}} \equiv \omega_{di}^2 \frac{\varepsilon_{opt}+2}{\varepsilon_{st}+2}, \qquad (40)$$

$$\Lambda_{st} = \frac{\omega_{pi}^2}{\omega_{di}^2}+\Lambda_{opt}, \quad \varepsilon_{st} = \frac{1+\frac{2}{3}\Lambda_{st}}{1-\frac{1}{3}\Lambda_{st}}$$

$$\omega_{Te(\pm)}^2 = \frac{1}{2}\left[\omega_{he+}^2 + \omega_{he-}^2 \pm \sqrt{\left(\omega_{he+}^2 + \omega_{he-}^2\right)^2 + \frac{4}{9}\omega_{pe+}^2\omega_{pe-}^2}\right],$$

where

$$\omega_{he(\pm)}^2 = \omega_{de(\pm)}^2 - \frac{1}{3}\omega_{pe(\pm)}^2.$$

From the obvious requirement $\varepsilon_{st} > 1$ from equality (18) it follows that for ion crystal dielectrics the condition on the parameter value $3 > \Lambda_{st} > 1$ is always satisfied. We also note that the frequency of transverse optical phonons (40) tends to zero at $\Lambda_{st} \to 3$, and the static dielectric constant increases indefinitely $\varepsilon_{st} \to \infty$ (the phenomenon of "polarization catastrophe" [18]). Under these conditions, a dielectric transforms into a ferroelectric state.

In the frequency range

$$\omega << \omega_{Li} \qquad (41)$$

the dielectric constant of an ion crystal is frequency independent and has constant value $\varepsilon = \varepsilon_{st}$, where

$$\varepsilon_{st} = \frac{\omega_{Li}^2}{\omega_{Ti}^2}\frac{\omega_{Le-}^2}{\omega_{Te-}^2}\frac{\omega_{Le+}^2}{\omega_{Te+}^2} \qquad (42)$$

is the static dielectric constant. On the other hand in the optical frequency range

$$\omega_{Te-}^2 >> \omega^2 >> \omega_{Li}^2 \qquad (43)$$

dielectric permeability is also constant $\varepsilon = \varepsilon_{opt}$

$$\varepsilon_{opt} = \frac{\omega_{Le-}^2}{\omega_{Te-}^2}\frac{\omega_{Le+}^2}{\omega_{Te+}^2} \qquad (44)$$

and for all ion dielectrics always $\varepsilon_{st} > \varepsilon_{opt}$. We note that from the expressions (42) and (44) imply the well-known Liddane-Sachs-Teller relation [15,16,18]

$$\frac{\omega_{Li}^2}{\omega_{Ti}^2} = \frac{\varepsilon_{st}}{\varepsilon_{opt}}. \qquad (45)$$

Formula (45) relates the ratio of the frequencies of longitudinal and transverse optical phonons with the values of static and optical dielectric constants. From inequality $\varepsilon_{st} > \varepsilon_{opt}$ important conclusion follows. Since Cherenkov radiation appears for a laser pulse when the condition

$$\frac{v_g^2}{c^2}\varepsilon_{st} > 1$$

is satisfied and the group velocity of the laser pulse in the optical range is equal $v_g = 1/\sqrt{\varepsilon_{opt}}$, then in the ion crystal the condition for the appearance of Cherenkov radiation in the microwave and terahertz ranges is always fulfilled.

The expression for the dielectric constant of the ion dielectric (36) can be given the usual and comfortable look

$$\varepsilon(\omega) = 1 - \frac{\Omega_{pi}^2}{\omega^2 - \omega_{Ti}^2} - \frac{\Omega_{pe-}^2}{\omega^2 - \omega_{Te-}^2} - \frac{\Omega_{pe+}^2}{\omega^2 - \omega_{Te+}^2}.$$

Here, the plasma frequencies are defined as follows

$$\Omega_{pi}^2 = (\varepsilon_{st} - \varepsilon_{opt})\omega_{Ti}^2,$$

$$\Omega_{pe+}^2 = \frac{\omega_{Te+}^2}{\omega_{Te+}^2 - \omega_{Te-}^2}\left[\omega_{Le}^2 - \omega_{Te}^2 - \omega_{Te-}^2(\varepsilon_{opt} - 1)\right],$$

$$\Omega_{pe-}^2 = \frac{\omega_{Te-}^2}{\omega_{Te+}^2 - \omega_{Te-}^2}\left[\omega_{Te+}^2(\varepsilon_{opt} - 1) - \omega_{Le}^2 + \omega_{Te}^2\right],$$

$$\omega_{Le}^2 = \omega_{Le+}^2 + \omega_{Le-}^2, \quad \omega_{Te}^2 = \omega_{Te+}^2 + \omega_{Te-}^2.$$

Fig. 2 shows the qualitative dependence of the dielectric constant on frequency, described by formula (36).

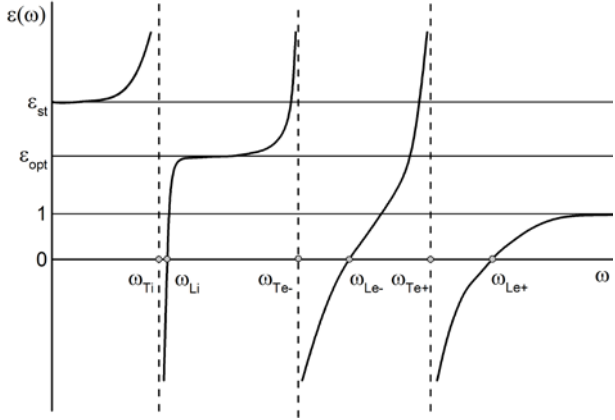

*Fig.2. Dependence of the dielectric constant on frequency*

In the frequency range $\omega^2 \ll \omega_{Ti}^2$, $\omega_{Li}^2 \ll \omega^2 \ll \omega_{Te-}^2$ dielectric constant is slightly dependent on frequency and equals $\varepsilon = \varepsilon_{st}$ and $\varepsilon = \varepsilon_{opt}$, correspondingly.

As an example, we give the measured values of the characteristic frequencies and dielectric permittivities for two alkali-halide crystals [15]:

NaCl

$\omega_{Ti} = 2\pi \cdot 4.86 10^{12},\quad \Omega_{pi} = 2\pi \cdot 6.9 10^{12},$

$\omega_{Te-} = 2\pi \cdot 1.89 10^{15},\quad \Omega_{pe-} = 2\pi \cdot 0.986 10^{15},$

$\omega_{Te+} = 2\pi \cdot 2.765 10^{15},\quad \Omega_{pe+} = 2\pi \cdot 2.77 10^{15},$

KCl

$\omega_{Ti} = 2\pi \cdot 4.27 10^{12},\quad \Omega_{pi} = 2\pi \cdot 6.785 10^{12},$

$\omega_{Te-} = 2\pi \cdot 1.85 10^{15},\quad \Omega_{pe-} = 2\pi \cdot 0.89 10^{15},$

$\omega_{Te+} = 2\pi \cdot 2.77 10^{15},\quad \Omega_{pe+} = 2\pi \cdot 2.46 10^{15}.$

Table 1 presents the values of static and optical permittivities, and also the frequencies of the optical transverse phonons for crystals of the alkali halide group [15]. Note that from these data it is not difficult to determine the frequencies $\omega_{Li}$ of longitudinal optical phonons from the Liddane-Sachs-Teller formula (38).

Table 1

| crystals | $\omega_{Ti} \cdot 10^{13} s^{-1}$ | $\lambda_{Ti}\ \mu m$ | $\varepsilon_{st}$ | $\varepsilon_{opt}$ |
|---|---|---|---|---|
| LiF | 5.78 | 32.6 | 9.27 | 1,92 |
| NaF | 4.64 | 7.4 | 6.0 | 1.74 |
| NaCl | 3.09 | 61.1 | 5.62 | 2.25 |
| NaI | 2.20 | 85.5 | 6.60 | 2.91 |
| KCl | 2.67 | 70.7 | 4.68 | 2.13 |
| KI | 1.85 | 102.0 | 4.94 | 2.69 |

## 2.2. DISPERSION PROPERTIES OF ION DIELECTRIC WAVEGUIDE

Let us now briefly discuss the question of the propagation of electromagnetic waves in an ion dielectric waveguide. Dispersion equations for potential longitudinal oscillations and electromagnetic waves have the form

$$\varepsilon(\omega) = 0, \quad (46)$$

$$\frac{\omega^2}{c^2}\varepsilon(\omega) - k_z^2 - \frac{\lambda_n^2}{b^2} = 0, \quad (47)$$

$k_z$ is longitudinal wave number. The dielectric constant is described by the formula (36).

The roots of dispersion equation (46) for longitudinal polarization oscillations determine the frequency of longitudinal optical phonons $\omega = \omega_{Li}$ and two frequencies $\omega = \omega_{Le(\pm)}$ of purely electron polarization oscillations of an ion crystal. The frequency of longitudinal optical phonons is in the infrared frequency range, and the frequencies of electron polarization oscillations belong to the optical or ultraviolet ranges.

Taking into account the expression for permittivity (36), the dispersion equation for transverse electromagnetic waves takes the form

$$\frac{\omega^2(\omega^2 - \omega_{Li}^2)(\omega^2 - \omega_{Le-}^2)(\omega^2 - \omega_{Le+}^2)}{(\omega^2 - \omega_{Ti}^2)(\omega^2 - \omega_{Te-}^2)(\omega^2 - \omega_{Te+}^2)} = \omega_n^2 + k_z^2 c^2, (48)$$

where $\omega_n = \frac{\lambda_n c}{b}$. First of all, we note that there are four frequency regions of opacity for electromagnetic waves $\omega < \omega_{ci-},\ \omega_{Ti} < \omega < \omega_{ci+},\ \omega_{Te-} < \omega < \omega_{ce-},\ \omega_{Te+} < \omega < \omega_{ce+}$, in which electromagnetic waves experience total internal reflection

$$k_z^2 = \frac{\omega^2}{c^2}\varepsilon(\omega) - \frac{\lambda_n^2}{b^2} < 0.$$

Cutoff frequencies $\omega = \omega_{ci,e(\pm)}$ can be determined from the dispersion equation (48) if $k_z = 0$

$$\frac{\omega^2(\omega^2 - \omega_{Li}^2)(\omega^2 - \omega_{Le-}^2)(\omega^2 - \omega_{Le+}^2)}{(\omega^2 - \omega_{Ti}^2)(\omega^2 - \omega_{Te-}^2)(\omega^2 - \omega_{Te+}^2)} = \omega_n^2. \quad (49)$$

For the infrared frequency range $\omega^2 \ll \omega_{Ti}^2$, instead of (49), we obtain the biquadratic equation

$$\frac{\omega^2(\omega^2 - \omega_{Li}^2)}{(\omega^2 - \omega_{Ti}^2)} = \frac{\omega_n^2}{\varepsilon_{opt}},$$

from which we find low-frequency cutoff frequencies

$$\omega_{ci(\mp)}^2 = \frac{1}{2}\left[\omega_{Li}^2 + \frac{\omega_n^2}{\varepsilon_{opt}} \mp \sqrt{\left(\omega_{Li}^2 + \frac{\omega_n^2}{\varepsilon_{opt}}\right)^2 - 4\omega_{Li}^2\frac{\omega_n^2}{\varepsilon_{opt}}}\right].$$

In the limit case $\omega_{Li}^2 \gg \omega_n^2/\varepsilon_{st}$, these expressions are simplified

$$\omega_{ci-}^2 = \frac{\omega_n^2}{\varepsilon_{st}}, \qquad \omega_{ci+}^2 = \omega_{Li}^2 + \frac{1}{\varepsilon_{eff}}\omega_n^2, \qquad (50)$$

$$\varepsilon_{eff} = \frac{\varepsilon_{st}\varepsilon_{opt}}{\varepsilon_{st} - \varepsilon_{opt}}.$$

The cutoff frequency $\omega_{ci+}$ slightly exceeds the frequency of longitudinal optical phonons. Accordingly, the first two frequency zones of propagation of electromagnetic waves are determined by the inequalities

$$\omega_{Ti} > \omega_1(k_z) > \omega_{ci-}, \quad \omega_{Te-} > \omega_2(k_z) > \omega_{ci+}$$

In the optical frequency range $\omega^2 \gg \omega_{Li}^2$, the algebraic equation for determining cutoff frequencies has the form

$$\frac{\omega^2(\omega^2 - \omega_{Le-}^2)(\omega^2 - \omega_{Le+}^2)}{(\omega^2 - \omega_{Te-}^2)(\omega^2 - \omega_{Te+}^2)} = \omega_n^2.$$

Assuming that $\omega_n^2$ is small parameter, the roots of this equation are easily found approximately

$$\omega_{ce-}^2 = \omega_{Le-}^2 + \frac{\omega_n^2}{\omega_{Le-}^2}\frac{\left(\omega_{Le-}^2 - \omega_{Te-}^2\right)\left(\omega_{Te+}^2 - \omega_{Le-}^2\right)}{\left(\omega_{Le+}^2 - \omega_{Le-}^2\right)},$$

$$\omega_{ce+}^2 = \omega_{Le+}^2 + \frac{\omega_n^2}{\omega_{Le+}^2}\frac{\left(\omega_{Le+}^2 - \omega_{Te+}^2\right)\left(\omega_{Le+}^2 - \omega_{Te-}^2\right)}{\left(\omega_{Le+}^2 - \omega_{Le-}^2\right)}.$$

The propagation zones of electromagnetic waves in the optical frequency range are determined by the inequalities

$$\omega_{Te+} > \omega_3(k_z) > \omega_{ce-}, \quad \omega_4(k_z) > \omega_{ce+}.$$

Let us pass to the analytical study of dispersion equation (48). In the infrared and microwave frequency ranges, dispersion equation (48) is simplified and takes the form

$$\frac{\omega^2(\omega^2 - \omega_{Li}^2)}{\omega^2 - \omega_{Ti}^2} = \frac{1}{\varepsilon_{opt}}(\omega_n^2 + k_z^2 c^2) \equiv K_n^2.$$

The roots of this equation are easy to find.

$$\omega_{1,2}^2(k_z) = \frac{1}{2}\left[\omega_{Li}^2 + K_n^2 \mp \sqrt{\left(\omega_{Li}^2 + K_n^2\right)^2 - 4\omega_{Ti}^2 K_n^2}\right].$$

In the region of small $K_n^2 \ll \omega_{Ti}^2$, instead of these expressions, we have

$$\omega_1^2(k_z) = \omega_{ci-}^2 + \frac{c^2}{\varepsilon_{st}}k_z^2, \quad \omega_2^2(k_z) = \omega_{ci+}^2 + \frac{c^2}{\varepsilon_{eff}}k_z^2. \quad (51)$$

For $k_z = 0$ from these expressions, low-frequency cutoff frequencies follow (50). The parabolic part of the ion branch $\omega_1(k_z)$ in the vicinity of the cutoff frequency is replaced by a linear plot $\omega_1(k_z) = k_z c/\sqrt{\varepsilon_{st}}$ at $k_z \gg \omega_{ce-}\sqrt{\varepsilon_{st}}/c$. As for the branch $\omega_2(k_z)$, its low-frequency (ion) part of the dispersion curve is described by a parabolic law. In the opposite (short-wave) limiting case $\omega_{Te-}^2 \gg K_n^2 \gg \omega_{Ti}^2$, the dispersion of these two branches of electromagnetic waves is described by the expressions

$$\omega_1^2(k_z) = \omega_{Ti}^2\left(1 - \frac{\omega_{Ti}^2 \varepsilon_{st}}{k_z^2 c^2}\right), \qquad (52)$$

$$\omega_2^2(k_z) = \frac{k_z^2 c^2}{\varepsilon_{opt}}\left(1 + \frac{\omega_{Ti}^2(\varepsilon_{st} - \varepsilon_{opt})}{k_z^2 c^2}\right). \qquad (53)$$

From these expressions it follows that the branch of oscillations $\omega_1(k_z)$ after the linear part of the dispersion $\omega_1(k_z) = k_z c/\sqrt{\varepsilon_{st}}$ asymptotically approaches from below to the frequency of transverse optical phonons $\omega_{Ti}$. For its part, the branch $\omega_2(k_z)$ from above approaches to the linear portion of the dispersion curve

$$\omega_2(k_z) = \frac{k_z c}{\sqrt{\varepsilon_{opt}}}.$$

In this frequency range, electromagnetic waves propagate with a phase velocity $v_{ph} = c/\sqrt{\varepsilon_{opt}}$ that exceeds the phase velocity of waves belonging to a branch $\omega_1(k_z)$ in its linear region, as always $\varepsilon_{st} > \varepsilon_{opt}$. Note that in the linear part of the dispersion curve, the branch $\omega_2(k_z)$ is mainly electronic, since in this frequency range the contribution of ions to the polarization of the dielectric, due to their large mass, is small.

Let us now consider the dispersion of electron electromagnetic waves in the optical range $\omega^2 \gg \omega_{Li}^2$. The dispersion equation describing the purely electronic branches of electromagnetic waves follows from (48) and has the form

$$\frac{\omega^2(\omega^2 - \omega_{Le-}^2)(\omega^2 - \omega_{Le+}^2)}{(\omega^2 - \omega_{Te-}^2)(\omega^2 - \omega_{Te+}^2)} = \omega_n^2 + k_z^2 c^2 \equiv k_n^2. \quad (54)$$

First of all, we consider a relatively low-frequency region of the optical frequency range $\omega_{Te-}^2 \gg \omega^2 \gg \omega_{Li}^2$. For this frequency range, from (54) we obtain

$$\omega_2^2(k_z) = \frac{k_z^2 c^2}{\varepsilon_{opt}}\left[1 - k_z^2 c^2\frac{\varepsilon_{opt}\omega_{Te}^2 - \omega_{Le}^2}{\omega_{Le-}^2\omega_{Le+}^2}\right].$$

From this expression it follows that the dispersion law, as in the case of (53), is close to linear, with the only difference being that with an increase of the longitudinal wave number, the dispersion curve $\omega_2(k_z)$ is shifted downward relative to the straight line $\omega = k_z c/\sqrt{\varepsilon_{opt}}$.

In the vicinity of the electronic cutoff frequencies, the dispersion curves look like parabolas

$$\omega_3^2(k_z) = \omega_{Le-}^2 + k_z^2\frac{(\omega_{Le-}^2 - \omega_{Te-}^2)(\omega_{Le-}^2 - \omega_{Te+}^2)}{(\omega_{Le-}^2 - \omega_{Le+}^2)\omega_{Le-}^2},$$

$$\omega_4^2(k_z) = \omega_{Le+}^2 + k_z^2\frac{(\omega_{Le+}^2 - \omega_{Te-}^2)(\omega_{Le+}^2 - \omega_{Te+}^2)}{(\omega_{Le+}^2 - \omega_{Le-}^2)\omega_{Le+}^2}.$$

And finally, in the short-wave limiting case $k_z c \gg \omega_{Le+}$ for electron electromagnetic branches $\omega_{2,3,4}(k_z)$, we have

$$\omega_2(k_z) = \omega_{Te-} - \frac{\omega_{Te-}}{2k_z^2 c^2}\frac{(\omega_{Le-}^2 - \omega_{Te-}^2)(\omega_{Le+}^2 - \omega_{Te-}^2)}{(\omega_{Te+}^2 - \omega_{Te-}^2)},$$

$$\omega_3(k_z) = \omega_{Te+} - \frac{\omega_{Te+}}{2k_z^2 c^2} \frac{(\omega_{Te+}^2 - \omega_{Le-}^2)(\omega_{Le+}^2 - \omega_{Te-}^2)}{(\omega_{Te+}^2 - \omega_{Te-}^2)},$$

$$\omega_4(k_z) = k_z c + \frac{1}{2}\frac{(\omega_{Le}^2 - \omega_{Te}^2)}{k_z c}.$$

From these expressions it follows that when $k_z \to \infty$ the branches $\omega_{2,3}(k_z)$ approaches from below to its limiting frequencies $\omega_{Te(\mp)}$, and the branch of the highest frequency $\omega_4(k_z)$ asymptotically approaches above to the "vacuum" dispersion line $\omega = k_z c$.

Fig. 3 shows the qualitative dependences of the frequency on the longitudinal wave number $k_z$ for radial harmonic with number $n$. In total, there are three branches of longitudinal oscillations $\omega = \omega_{Li}, \omega_{Le(\pm)}$ (blue lines) and four branches $\omega_\alpha(k_z)$ ($\alpha = 1 \div 4$) of electromagnetic waves (red curves).

The low-frequency branch corresponds to the longitudinal optical phonons, and the other two branches are polarization electron oscillations.

As for the electromagnetic branches (red curves), the lowest frequency (ion) branch $\omega_1(k_z)$ is in the infrared and microwave ranges $\omega_{Ti} > \omega_1(k_z) > \omega_{ci-}$. In the frequency range $\omega_{Ti} \gg \omega \gg \omega_{ci-}$, the dispersion curve has a linear plot $\omega_1(k_z) = k_z c / \sqrt{\varepsilon_{st}}$.

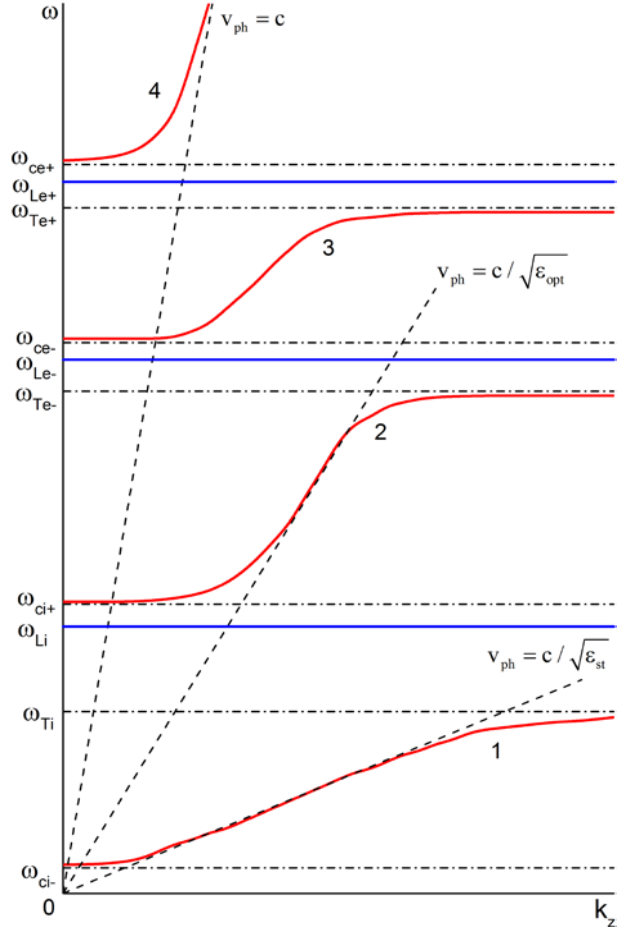

*Fig. 3. Dispersion curves of longitudinal oscillations (blue) and electromagnetic waves (red) on the plane ($\omega, k_z$) for ion dielectric*

The frequencies of the electromagnetic branch 2 are within $\omega_{Te-ci} > \omega_2(k_z) > \omega_+$. The low-frequency section of this branch corresponds to the infrared frequency range and the high-frequency region corresponds to the optical one. This branch also has a linear dispersion region $\omega_2(k_z)_{opt} = k_z c / \sqrt{\varepsilon}$. The tilt angle of this line exceeds the tilt angle of the straight section of branch $\omega_1(k_z)$. And finally, branches $\omega_{3,4}(k_z)$ are purely electron branches and locate in the optical and ultraviolet frequency ranges. The phase velocity of electromagnetic waves belonging to the fourth branch exceeds the speed of light and in the limiting case approaches it.

## 2.3. CALCULATION OF GREEN'S FUNCTION

The Green function (35), contains two terms that describe the excitation of longitudinal potential oscillations and electromagnetic waves. The potential Green's function $G_l(r, \tau - \tau_0)$ has only simple poles, which are the zeros of the dielectric constant $\varepsilon(\omega) = 0$. The frequency spectrum of longitudinal oscillations contains the frequency of longitudinal optical phonons $\omega_{Li}$ and the frequencies $\omega_{Le(\mp)}$ of electron polarization oscillations. Below we restrict ourselves to the study of wake fields in the infrared and lower frequency ranges. This is due to the fact that for effective wake field excitation by a laser pulse necessary to achieve coherency of excitation. For this, it is necessary that the longitudinal and transverse dimensions of the laser pulse be smaller (substantially less) than the length of the radiated wave. For the optical and especially the ultraviolet frequency ranges, this requirement is very problematic. And if this requirement is not satisfied, the amplitude of the wake waves will be negligible.

Calculating the residues in the integral $G_l(r, \tau - \tau_0)$ at the poles $\omega = \pm \omega_{Li} - i0$, we find the following expression for the potential Green function

$$G_l = \frac{4\pi}{3}\frac{\omega_{Li}^2}{\varepsilon_{eff}}\Gamma_d \Phi_i(r) \vartheta(\tau - \tau_0) \cos\omega_{Li}(\tau - \tau_0), \quad (55)$$

where $\vartheta(\tau - \tau_0)$ is the Heaviside function,

$$\Phi_i(r) = R(r) + \frac{\varepsilon_L - 1}{6} k_i^2 \int_0^b G_r(r, r_0) R(r_0) r_0 dr_0,$$

$$G_r(r, r_0) = \frac{1}{I_0(k_i b)} \begin{cases} I_0(k_i r)\Delta_0(k_i r_0), & r < r_0, \\ I_0(k_i r_0)\Delta_0(k_i r), & r > r_0, \end{cases}$$

$$\Delta_0(k_i r) = I_0(k_i r) K_0(k_i b) - I_0(k_i b) K_0(k_i r),$$

$$k_i = \omega_{Li}/v_g, \quad \Gamma_d = \frac{\alpha_L^{(+)}}{\omega_{de+}^2} + \frac{\alpha_L^{(-)}}{\omega_{de-}^2}.$$

Term in the total Green's function (35) $G_{tr}(r, \tau - \tau_0)$ describes the Cherenkov excitation of the eigen electromagnetic waves of the dielectric waveguide. Integrands of Fourier integrals $S_n(\tau - \tau_0)$ contain only simple poles, which are the roots of the equation

$$\Delta_n(\omega) = 0. \quad (56)$$

These roots determine the frequency spectrum of electromagnetic radial harmonics of a dielectric waveguide, which are in Cherenkov synchronism with a laser pulse. For a qualitative analysis of the location of these poles on the complex plane $\omega$, the spectrum equation (56) is conveniently represented as

$$\frac{(\omega^2 - \omega_{Li}^2)(\omega^2 - \omega_{Le-}^2)(\omega^2 - \omega_{Le+}^2)}{(\omega^2 - \omega_{Ti}^2)(\omega^2 - \omega_{Te-}^2)(\omega^2 - \omega_{Te+}^2)} = $$
$$= \frac{1}{\beta_g^2} + \frac{\omega_n^2}{\omega^2} \equiv f(\omega) \qquad (57)$$

$\beta_g = v_g / c$. A graphical solution of this equation shows that there are always three positive roots $\omega^2 = \omega_{ni}^2, \omega^2 = \omega_{ne(\mp)}^2$ and one negative root $\omega^2 = -v_{nst}^2$. The positive roots correspond to the frequencies of the eigen electromagnetic waves of the dielectric waveguide, which are in Cherenkov synchronism with the laser pulse $\omega = k_n(\omega)v_g$. The negative root (imaginary frequency) describes a quasistatic electromagnetic field that is localized in the region of the laser pulse and exponentially decreases with distance from it. The lowest frequency $\omega_{ni}$ belongs to the infrared branch (see Fig. 3). The frequency $\omega_{ne-}$ belongs to a hybrid (ion-electron) branch. And finally, the third highest frequency $\omega_{ne+}$ belongs to the purely electronic optical branch of electromagnetic waves $\omega_3(k_z)$. As we are interested in the infrared (microwave) frequency range, the spectrum equation (57) is simplified and takes the form

$$\omega^2 \frac{\omega^2 - \omega_{Li}^2}{\omega^2 - \omega_{Ti}^2} = \frac{\omega^2}{\beta_g^2 \varepsilon_{opt}} + \frac{\omega_n^2}{\varepsilon_{opt}}.$$

The roots of this equation are found easily.

$$\omega_{ni(\mp)}^2 = \frac{1}{2\Delta_{opt}}\left(\omega_{Ti}^2 \Delta_{st} + \omega_n^2\right) \mp $$
$$\mp \sqrt{\frac{1}{4\Delta_{opt}^2}\left(\omega_{Ti}^2 \Delta_{st} + \omega_n^2\right)^2 + \frac{\omega_n^2 \omega_{Ti}^2}{\Delta_{opt}}}. \qquad (58)$$

Here $\Delta_{st} = \dfrac{\beta_g^2 \varepsilon_{st} - 1}{\beta_g^2}$, $\Delta_{opt} = \dfrac{1 - \beta_g^2 \varepsilon_{opt}}{\beta_g^2}$.

When the condition is fulfilled

$$\frac{4\omega_n^2 \omega_{Ti}^2 |\Delta_{opt}|}{\left(\omega_{Ti}^2 \Delta_{st} + \omega_n^2\right)^2} \ll 1, \qquad (59)$$

the expressions for squares of frequencies (58) are simplified

$$\omega_{ni-} = \pm \omega_{nlf} - i0, \quad \omega_{nlf} = \sqrt{\frac{\omega_n^2 \beta_g^2}{\beta_g^2 \varepsilon_{st} - 1 + \beta_g^2 \frac{\omega_n^2}{\omega_{Ti}^2}}}, \qquad (60)$$

$$\omega_{ni+} = \pm \omega_{nhf} - i0, \quad \omega_{nhf} = \sqrt{\frac{\omega_{Ti}^2\left(\beta_g^2 \varepsilon_{st} - 1\right) + \omega_n^2 \beta_g^2}{\beta_g^2 \varepsilon_{opt} - 1}}. \qquad (61)$$

Taking into account the expressions for frequencies (60), (61), inequality (59) can be rewritten as follows

$$\frac{4\omega_{ni-}^2}{\omega_{ni+}^2} \ll 1.$$

The frequency $\omega_{ni-}$ belongs to the ionic branch 1 (see Fig. 3), and the frequency $\omega_{ni+}$ belongs to the hybrid ion-electron branch 2. The frequency $\omega_{ni+}$ formally tends to infinity $\omega_{ni+} \to \infty$ at $\beta_g \to 1/\sqrt{\varepsilon_{opt}}$. Physically, this means that in such a situation, the frequency $\omega_{ni+}$ shifts to the more high-frequency electronic region of the hybrid ion-electron branch 2. Since in the optical frequency range $\omega_{Te-} \gg \omega \gg \omega_{Li}$ $\beta_g \approx 1/\sqrt{\varepsilon_{opt}}$ for determination of the frequency (61) under such conditions, the dielectric constant $\varepsilon(\omega)$ in this frequency range can be represented as

$$\varepsilon(\omega) = \varepsilon_{opt}\left[1 - \frac{\omega_{Li}^2 - \omega_{Ti}^2}{\omega^2} + \omega^2\left(\frac{1}{\bar{\omega}_{Te}^2} - \frac{1}{\bar{\omega}_{Le}^2}\right)\right], \qquad (62)$$

where

$$\bar{\omega}_{Te}^2 = \frac{\omega_{Te-}^2 \omega_{Te+}^2}{\omega_{Te-}^2 + \omega_{Te+}^2}, \quad \bar{\omega}_{Le}^2 = \frac{\omega_{Le-}^2 \omega_{Le+}^2}{\omega_{Le-}^2 + \omega_{Le+}^2}$$

are reduced frequencies. Given this expression for permittivity, the spectrum function $\Delta_n(\omega)$ can be written as follows

$$\Delta_n(\omega) = \frac{1}{c^2 \omega_{TLe}^2}\left(\omega^4 - \Delta_{opt} \omega_{TLe}^2 \omega^2 - \omega_{\Gamma n}^4\right). \qquad (63)$$

Here

$$\omega_{TLe}^2 = \frac{\omega_{Te-}^2 \omega_{Te+}^2}{\omega_{Te}^2 \varepsilon_{opt} - \omega_{Le}^2}, \quad \omega_{\Gamma n}^4 = \omega_{TLe}^2 \Omega_{nTi}^2,$$

$$\Omega_{nTi}^2 = \omega_{Ti}^2\left(\varepsilon_{st} - \varepsilon_{opt}\right) + \omega_n^2, \quad \omega_n = \frac{\lambda_n c}{b},$$

$\Delta_{opt}$ is detuning between the group velocity of the laser pulse and the phase velocity of electromagnetic waves $v_{ph} = 1/\sqrt{\varepsilon_{opt}}$ in the optical and infrared frequency ranges $\omega_{Te-}^2 \gg \omega^2 \gg \omega_{Li}^2$. For the group velocity of the eigen electromagnetic waves of a dielectric waveguide, we have the following expression

$$\beta_g^2 = \frac{\varepsilon(\omega^2) - \dfrac{\omega_n^2}{\omega^2}}{\left[\dfrac{\partial}{\partial \omega^2}\omega^2 \varepsilon(\omega^2)\right]^2}.$$

Using the approximate expression for the dielectric constant (62) in the optical frequency range $\omega_{Te-}^2 \gg \omega^2 \gg \omega_{Li}^2$, we find the following expression for the group velocity

$$\beta_g^2 = \frac{v_g^2}{c^2} = \frac{1}{\varepsilon_{opt}}\left[1 - \frac{1}{\varepsilon_{opt}}\left(\frac{\Omega_{nTi}^2}{\omega^2} + 3\frac{\omega^2}{\omega_{TLe}^2}\right)\right].$$

Therefore the detuning parameter

$$\Delta_{opt} = \frac{\Omega_{nTi}^2}{\omega_L^2} + 3\frac{\omega_L^2}{\omega_{TLe}^2}$$

is always positive. From the spectrum equation $\Delta_n(\omega) = 0$, taking into account (63), the biquadratic equation follows

$$\omega^4 - \Delta_{opt}\omega_{TLe}^2\omega^2 - \omega_{\Gamma n}^4 = 0 \qquad (64)$$

for determining the frequencies of electromagnetic waves synchronous with a laser pulse. A positive root relative to the square of the frequency corresponds to a pair of frequencies (poles) located on the real axis

$$\omega = \pm \omega_{nhf} - i0, \qquad (65)$$

$$\omega_{nhf} = \sqrt{\frac{1}{2}\left(\omega_{LTe}^2 \Delta_{opt} + \sqrt{\omega_{LTe}^4 \Delta_{opt}^2 + 4\omega_{\Gamma n}^4}\right)}.$$

The negative root of the biquadratic equation (64) corresponds to a pair of frequencies (poles) located on the imaginary axis

$$\omega = \pm i\nu_{nst}, \qquad (66)$$

$$\nu_{nst} = \sqrt{\frac{1}{2}\left(\sqrt{\omega_{LTe}^4 \Delta_{opt}^2 + 4\omega_{\Gamma n}^4} - \omega_{LTe}^2 \Delta_{opt}\right)}.$$

If the detuning is small

$$\frac{\omega_{LTe}^4 \Delta_{opt}^2}{4\omega_{\Gamma n}^4} = \frac{\omega_{LTe}^2 \Delta_{opt}^2}{4\Omega_{Ti}^2} \ll 1,$$

then the values modules of real and imaginary roots coincide

$$\omega_{nhf} = \nu_{nst} = \omega_{\Gamma n}$$

In this case, the frequency has a hybrid ion-electron nature. For large values of the detuning parameter

$$\frac{\omega_{LTe}^2 \Delta_{opt}^2}{4\Omega_{Ti}^2} \gg 1 \qquad (67)$$

values $\omega_{nhf}$ and $\nu_{nst}$ are very different $\omega_{nhf}^2 \gg \nu_{nst}^2$,

$$\omega_{nhf}^2 = \omega_{LTe}^2 \Delta_{opt} = \omega_{LTe}^2 \frac{1-\beta_g^2 \varepsilon_{opt}}{\beta_g^2}, \qquad (68)$$

$$\nu_{nst}^2 = \frac{\Omega_{nLi}^2}{\Delta_{opt}} = \Omega_{nLi}^2 \frac{\beta_g^2}{1-\beta_g^2 \varepsilon_{opt}}.$$

Frequency (68) is purely electronic and belongs to the electronic region of the ion-electron branch 2.

Calculating the residues in poles (60), (65), (66) we find the expression for the Green function

$$G_{tr} = -\frac{\pi}{3} e_{st} \Gamma_d \vartheta(\tau - \tau_0) \sum_{n=1}^{\infty} \omega_{nlf}^2 \sigma_n J_0\left(\lambda_n \frac{r}{b}\right) \cos \omega_{nlf}(\tau - \tau_0) -$$

$$-\frac{\pi}{3} e_{opt} \Gamma_d \sum_{n=1}^{\infty} \omega_{Gn}^2 \sigma_n J_0\left(\lambda_n \frac{r}{b}\right) \Big[\vartheta(\tau - \tau_0) \cos \omega_{nhf}(\tau - \tau_0) -$$

$$-\frac{1}{2} sign(\tau - \tau_0) e^{-\nu_{nst}|\tau - \tau_0|}\Big], \qquad (69)$$

where

$$\sigma_n = \frac{\rho_n}{N_n}, \quad e_{st} = \frac{(\varepsilon_{opt}-1)}{\varepsilon_{st}} \frac{(\varepsilon_{st}+2)}{3},$$

$$e_{opt} = \frac{(\varepsilon_{opt}-1)}{\varepsilon_{opt}} \frac{(\varepsilon_{opt}+2)}{3}, \quad \omega_{Gn}^2 = \frac{\omega_{TLe}^2 \omega_n^2}{\omega_{nhf}^2 + \nu_{nst}^2}.$$

The first term in the expression for the electromagnetic Green's function (69) describes the electric field in the microwave (terahertz) frequency range $\omega_{Li}^2 \gg \omega_{nlf}^2$ (ion branch 1) and is a set of eigen electromagnetic waves with frequencies $\omega_{nlf}$. The second term in expression (69) describes electromagnetic field which belongs to branch 2 in the infrared frequency range. The longitudinal structure of this field is more complicated. Each radial harmonic contains a wake monochromatic wave, as well as a bipolar antisymmetric solitary pulse. Moreover, the height of this pulse is exactly two times smaller than the amplitude of the wake wave. The characteristic width of the polarization pulse is equal to $\Delta \tau = 1/\nu_{nst}$.

## 2.4. THE EXCITATION OF WAKE FIELD BY LASER PULSE

The wakefield excited by a laser pulse is described by convolution (33), in which the Green function is the key element. We first consider the excitation of longitudinal optical phonons. Using the potential polarization part of the Green function (55), we obtain the following expression for the wake field of longitudinal optical phonons

$$E_{iz}(r,\tau) = E_{Li} \Phi_i(r) Z(\omega_{Li}\tau), \qquad (70)$$

where

$$Z(\omega \tau) = \frac{1}{t_L} \int_{-\infty}^{\tau} T(\tau_0/t_L) \cos \omega(\tau - \tau_0) d\tau_0,$$

$$E_{Li} = \frac{2\pi}{3} \frac{(\varepsilon_{st} - \varepsilon_{opt})}{\varepsilon_{st} \varepsilon_{opt}} \frac{(\varepsilon_{opt}+2)}{3} \frac{e\omega_{Li}^2}{v_g^2} \frac{v_g t_L}{r_{cl}} \omega_L^2 \kappa_d a_0^2, \quad (71)$$

$t_L$ is characteristic duration of a laser pulse, $\varepsilon_L \approx \varepsilon_{opt}$,

$$\kappa_d = \Gamma_d N_0, \quad r_{cl} = e^2/mc^2, \quad a_0^2 = \left(\frac{e}{mc\omega_L}\right)^2 I_0.$$

$$\Gamma_d = \frac{\alpha_L^{(+)}}{\omega_{de+}^2} + \frac{\alpha_L^{(-)}}{\omega_{de-}^2}.$$

The wake function $Z(\omega \tau)$ describes the distribution of the wakefield on frequency $\omega$ in the longitudinal direction at each moment of time. We will consider a laser pulse with a symmetric longitudinal profile $T(\tau_0/t_L) = T(-\tau_0/t_L)$. The wake function $Z(\omega \tau)$ is conveniently represented as

$$Z(\omega \tau) = \hat{T}(\Omega) \vartheta(\tau) \cos \omega \tau - X(\bar{\tau}), \qquad (72)$$

where $\Omega = \omega t_L$, $\bar{\tau} = \tau/t_L$,

$$X(\bar{\tau}) = sign\tau \int_{|\bar{\tau}|}^{\infty} T(s) \cos \Omega(|\bar{\tau}| - s) ds,$$

$$\hat{T}(\Omega) = 2 \int_0^{\infty} T(s) \cos(\Omega s) ds.$$

The first term in (72) describes the wake wave propagating behind the laser pulse. The amplitude of the wake wave is equal to the Fourier amplitude function $T(\tau_0/t_L)$, which describes the longitudinal profile of the laser pulse. The second term in (72) describes a bipolar antisymmetric pulse of a polarization field localized in the region of a laser pulse. The field of this pulse decreases and tends to zero with increasing distance from the laser pulse.

Behind a laser pulse, the wake field (70) of longitudinal optical phonons has the form of a monochromatic wave

$$E_{iz}(r,\tau) = E_{Li}\Phi_i(r)\hat{T}(\Omega_{Li})\cos\omega_{Li}\tau, \quad \Omega_{Li} = \omega_{Li}t_L.$$

Let us give expressions for the Fourier amplitude $\hat{T}(\Omega_{Li})$ for two model longitudinal profiles of a laser pulse: Gaussian and power ones

$$T(\tau_0/t_L) = e^{-\tau_0^2/t_L^2}, \quad \hat{T}(\Omega) = \sqrt{\pi}e^{-\Omega^2/4}, \quad (73)$$

$$T(\tau_0/t_L) = \frac{1}{1+\tau_0^2/t_L^2}, \quad \hat{T}(\Omega) = \pi e^{-\Omega}.$$

Longitudinal optical phonons are most efficiently radiated when the coherence condition $\omega_{Li}t_L \leq 1$ is satisfied. If the condition $\omega_{Li}t_L \gg 1$ is satisfied, then the longitudinal optical phonons are radiated incoherently and the amplitude of the wake wave is exponentially small.

We present the expressions for the amplitudes of longitudinal optical phonons $E_{Li}$. For two alkaline halide ion dielectrics: sodium chloride NaCl and potassium iodide KI, we have.

$$E_{Li}(NaCl) = 7.4 \cdot 10^4 \frac{N_L}{\lambda_L(\mu m)} a_0^2 \ (V/cm),$$

where $N_L$ is number of wavelengths in the laser pulse, $f_{Li} = 7.62 \cdot 10^{12} Hz$ is frequency of longitudinal optical phonons. In the case of potassium iodide, we obtain

$$E_{Li}(KI) = 1.16 \cdot 10^4 \frac{N_L}{\lambda_L(\mu m)} a_0^2 \ (V/cm),$$

$f_{Li} = 4 \cdot 10^{12} Hz$. The amplitude of the longitudinal optical phonons and their frequency are lower than in the case of potassium chloride.

For $\lambda_L = 1\mu m$, $N = 30$ and $a_0 = 1$ we find $E_{Li}(NaCl) = 2.2 MV/cm$, $E_{Li}(KI) = 0.35 MV/cm$.

Let us now consider the excitation of electromagnetic waves by a laser pulse. Taking into account the advantage of the electromagnetic Green's function, we obtain the wake electromagnetic field as a superposition of radial harmonics

$$E_{tz}(r,\tau) = -E_{lf}\sum_{n=1}^{\infty}\lambda_n^2\hat{\sigma}_n \frac{J_0\left(\lambda_n\frac{r}{b}\right)}{J_1^2(\lambda_n)}Z(\omega_{nlf}\tau) - $$

$$-E_{hf}\sum_{n=1}^{\infty}\frac{\omega_{TLe}^2\lambda_n^2\hat{\sigma}_n}{\omega_{nhf}^2+\nu_{nst}^2}\frac{J_0\left(\lambda_n\frac{r}{b}\right)}{J_1^2(\lambda_n)}\left[Z(\omega_{nhf}\tau) - \frac{1}{2}Y(\nu_{nst}\tau)\right],$$

where

$$Y(\nu_{nst}\tau) = \frac{1}{t_L}\int_{-\infty}^{\infty}T(\tau_0/t_L)sign(\tau-\tau_0)e^{-\nu_{nst}|\tau-\tau_0|}d\tau_0,$$

$$E_{lf} = \varepsilon_{lf}\omega_L^2\kappa_d\frac{\pi r_b^2}{b^2}\frac{v_g t_L}{r_{cl}}\frac{e}{b^2}a_0^2, \quad (74)$$

$$\varepsilon_{lf} = \frac{1}{3}\frac{\varepsilon_{opt}}{\varepsilon_{st}}\frac{\varepsilon_{opt}-1}{\varepsilon_{st}-\varepsilon_{opt}}\frac{\varepsilon_{st}+2}{3}\frac{\varepsilon_{opt}+2}{3},$$

$$E_{hf} = \varepsilon_{hf}\omega_L^2\kappa_d\frac{\pi r_b^2}{b^2}\frac{v_g t_L}{r_{cl}}\frac{e}{b^2}a_0^2, \quad (75)$$

$$\varepsilon_{hf} = \frac{1}{3}(\varepsilon_{opt}-1)\left(\frac{\varepsilon_{opt}+2}{3}\right)^2.$$

$$\hat{\sigma}_n = \int_0^{b/r_L} R(\rho)J_0\left(\lambda_n\frac{r_L}{b}\rho\right)\rho d\rho.$$

Expressions for quantities $E_{lf,hf}$ (74), (75) can be written in terms of the total energy of the laser pulse

$$W_L = \frac{c}{8\pi}I_0 s_L t_L \hat{\sigma}\hat{\tau}, \quad (76)$$

where $s_L = \pi r_L^2$ is the cross section area of the laser pulse

$$\hat{\sigma} = 2\int_0^{\infty}R(\rho)\rho d\rho, \quad \hat{\tau} = \int_{-\infty}^{\infty}T(\tau)d\tau.$$

The expression for the full energy of the laser pulse (76) can be converted to the form

$$W_L = \frac{c}{8\pi}I_0 s_L t_L \hat{\sigma}\hat{\tau},$$

As a result, instead of expressions (74), (75), we obtain

$$E_{lf,hf} = 8\pi\varepsilon_{lf,hf}\beta_g \frac{\kappa_d c^2}{b^2}\frac{1}{\hat{\sigma}\hat{\tau}}\frac{Q_{Leff}}{b^2}, \quad (77)$$

where $Q_{Leff} = e\frac{W_L}{mc^2}$ is the effective "electric charge" of the laser pulse.

Behind the laser pulse $\tau/t_L \gg 1, \omega_{nlf}\tau \gg 1$, $\omega_{nhf}\tau \gg 1$ the pulse fields are negligible and only the set of eigen waves of the dielectric waveguide remains

$$E_{tz}(r,\tau) = -\sum_{n=1}^{\infty}E_{nlf}J_0\left(\lambda_n\frac{r}{b}\right)\cos(\omega_{nlf}\tau) - $$

$$-\sum_{n=1}^{\infty}E_{nhf}J_0\left(\lambda_n\frac{r}{b}\right)\cos(\omega_{nhf}\tau), \quad (78)$$

$$\Omega_{nlf} = \omega_{nlf}t_L, \quad \Omega_{nhf} = \omega_{nhf}t_L,$$

$$E_{nlf} = E_{lf}\frac{\lambda_n^2}{J_1^2(\lambda_n)}\hat{\sigma}_n\hat{T}(\Omega_{nlf}),$$

$$E_{nhf} = E_{hf}\frac{\lambda_n^2}{J_1^2(\lambda_n)}\hat{\sigma}_n\frac{\omega_{TLe}^2}{\omega_{nhf}^2+\nu_{nst}^2}\hat{T}(\Omega_{nhf}),$$

Let us consider, for example, a laser pulse that has a Gaussian profile both in the longitudinal direction (73) and in the transverse one

$$R(r_0/r_L) = \exp(-r_0^2/r_L^2).$$

The radius of the laser pulse is small compared with the radius of the dielectric waveguide $r_L \ll b$. In this case,

for the expansion coefficients $\sigma_n$ in the series (74) we have

$$\hat{\sigma}_n = \exp\left(-\frac{\lambda_n^2 r_L^2}{4b^2}\right).$$

Accordingly, for the wake electromagnetic field instead of (78) we obtain

$$E_{tz}(r,\tau) = -\sqrt{\pi}\sum_{n=1}^{\infty}\lambda_n^2 \hat{\sigma}_n \frac{J_0\left(\lambda_n \frac{r}{b}\right)}{J_1^2(\lambda_n)}\left[E_{lf} e^{-\frac{\omega_{nlf}^2 t_L^2}{4}}\cos(\omega_{nlf}\tau) + \right.$$
$$\left. + E_{hf}\frac{\omega_{TLe}^2}{\omega_{nhf}^2 + \nu_{nst}^2} e^{-\frac{\omega_{nhf}^2 t_L^2}{4}}\cos(\omega_{nhf}\tau)\right]. \quad (79)$$

An important conclusion follows from formulas (77), (78), (79) that, for a short laser pulse $\omega_{lf}^2 t_L^2/4 \ll 1$, $\omega_{hf}^2 t_L^2/4 \ll 1$, the radial harmonic amplitudes for which the condition $\lambda_n^2 r_L^2/4b^2 \ll 1$ is satisfied do not depend on the sizes of the laser pulse and its frequency, but are determined only by the total energy of the laser pulse and the parameters of the dielectric waveguide.

If the laser pulse is long at the scale of the minimum period of electronic electromagnetic wave $\omega_{1hf} t_L > 1$, but short compared with the periods of ion electromagnetic waves $\omega_{nlf} t_L \ll 1$, then low-frequency ion electromagnetic waves will be most effectively excited. Under these conditions, only low-frequency waves are radiated coherently by a laser pulse.

To estimate the efficiency of excitation of a wake electromagnetic field in ion dielectrics of an alkali halide group, we use the data given in Tables 1, 2. However, we first transform the key parameter $\kappa_d$ included in the expressions for the amplitudes $E_{lf}$, $E_{hf}$ of the electromagnetic field (78), (79). We represent the electronic polarizabilities of ions $\alpha_L^{(\pm)}$ in the form

$$\alpha_L^{(\pm)} = \frac{\alpha_{st}^{(\pm)}}{1-\omega_L^2/\omega_{de(\pm)}^2},$$

where

$$\alpha_{st}^{(\pm)} = \frac{eq^{(\pm)}}{m\omega_{de(\pm)}^2} \quad (80)$$

is the electronic polarizability of ions in a static electric field. The values of these electron polarizabilities for alkali halide ions are given in Table 2 [15].

Table 2. *Electron polarizabilities of alkali and halide ions $\alpha_{st}^{(\pm)}$ in units $10^{-24} cm^3$*

|  | $Li^+$ | $Na^+$ | $K^+$ | $F^-$ | $Cl^-$ | $I^-$ |
|---|---|---|---|---|---|---|
| in crystal | 0.045 | 0.28 | 1.13 | 0.86 | 2.9 | 6.4 |
| free ions | 0.03 | 0.18 | 0.83 | 1.04 | 3.66 | 7.1 |

From relation (80) we find the relationship between the frequencies of dipole oscillations of ions and their electron polarizabilities

$$\omega_{de(\pm)}^2 = \frac{eq^{(\pm)}}{m\alpha_{st}^{(\pm)}}.$$

As a result, we obtain the following expression for the coefficient $\kappa_d$, convenient for numerical calculations

$$\kappa_d = \frac{m}{e^2}N_0\left(\frac{\alpha_{st}^{(-)^2}}{Z_-} + \frac{\alpha_{st}^{(+)^2}}{Z_+}\right),$$

where $Z_\pm$ are charge numbers of ions.

Below we consider two ion alkali halide crystals NaCl and KI. Their choice is due to the relatively high polarizability of ions (see Table 2). For these crystals, the amplitude $E_{lf}$ can be represented in a form convenient for numerical estimates

$$E_{lf}(NaCl) = 2.6\cdot 10^5 \frac{\eta_L N_L a_0^2}{\lambda_L(\mu m)[b(\mu m)]^2}\ (V/cm),$$

or $E_{lf}(NaCl) = 1.625\cdot 10^9 \frac{W(J)}{[b(\mu m)]^4}\frac{1}{\hat{\sigma}\hat{\tau}}\ (V/cm)$;

$$E_{lf}(KI) = 5.52\cdot 10^5 \frac{\eta_L N_L a_0^2}{\lambda_L(\mu m)[b(\mu m)]^2}\ (V/cm),$$

or $E_{lf}(KI) = 3.45\cdot 10^9 \frac{W(J)}{[b(\mu m)]^4}\frac{1}{\hat{\sigma}\hat{\tau}}\ (V/cm)$,

$\eta_L = r_L^2/b^2$, $N_L = ct_L/\lambda_L$ is number of wavelengths in a laser pulse. Accordingly for the amplitude of radial harmonic with the number $n$ of the total low frequence field (79) for a laser pulse with Gaussian longitudinal and transverse profiles, we have

$$E_{nlf} = \sqrt{\pi}E_{lf}\frac{\lambda_n^2}{J_1^2(\lambda_n)}e^{-\frac{\lambda_n^2 r_L^2}{4b^2} - \frac{\omega_{nfl}^2 t_L^2}{4}}.$$

The frequencies of the low-frequency radial harmonics for these crystals are equal

$$\omega_{nlf}(NaCl) = 2\pi\cdot\frac{0.36\lambda_n}{b(\mu m)}10^{14}\ s^{-1},$$

$$\omega_{nlf}(KI) = 2\pi\cdot\frac{0.32\lambda_n}{b(\mu m)}10^{14}\ s^{-1}.$$

For a KI crystal with the radius $b = 50\mu m$ amplitudes and frequencies of the first two radial harmonics, excited by a laser pulse with a Gaussian longitudinal and transverse profiles, are equal

$$f_{1lf} = 1.5 THz,\ E_{1lf} = 1,12\cdot 10^4 W(J) V/cm,$$
$$f_{2lf} = 3.5 THz,\ E_{2lf} = 1,45\cdot 10^5 W(J) V/cm.$$

These relations are valid for the case of coherent excitation of radial harmonics by a laser pulse.

We define the power of excited wake electromagnetic waves as the total energy flow carried by the corresponding wave along the dielectric waveguide

$$P = \frac{c}{4\pi}\int_0^b E_r H_\varphi 2\pi r dr,$$

where $E_r$ and $H_\varphi$ are the radial component of the electric field and the azimuthal component of the magnetic field of the corresponding wave. Then, using the expression for the wake field and integrating over the cross section of the waveguide, we obtain the following expression for the total power of electromagnetic radiation

$$P = \pi b^2 \left( \sum_{n=1}^{\infty} P_{nlf} + \sum_{n=1}^{\infty} P_{nhf} \right), \qquad (81)$$

$$P_{nlf} = P_{lf} \frac{\omega_{nlf}^2 b^2}{v_g^2} \lambda_n^2 J_1^2(\lambda_n) \sigma_n^2 \hat{T}^2(\Omega_{nlf}),$$

$$P_{nhf} = P_{hf} \frac{\omega_{nhf}^2 b^2}{v_g^2} \frac{\omega_{TLe}^4}{(\omega_{nhf}^2 + v_{nst}^2)^2} \lambda_n^2 J_1^2(\lambda_n) \sigma_n^2 \hat{T}^2(\Omega_{nlf}),$$

$$P_{lf} = \frac{1}{8\pi} v_g E_{lf}^2 \varepsilon_{st}, \quad P_{hf} = \frac{1}{8\pi} v_g E_{hf}^2 \varepsilon_{opt}.$$

The summation in (81) is performed according to the numbers of radial harmonics of the dielectric waveguide, $P_{nlf}$ is the average partial power density over the cross section of the waveguide of the eigen wave belonging to the low-frequency branch and $P_{nhf}$ is the average partial power of the more high frequency eigen wave belonging to the branch 2.

It should be noted that the wavelength of the electromagnetic wave $\lambda_{1hf} = 2\pi c/\omega_{1hf}$ belonging to branch 2 (see Fig. 3), for a dielectric waveguide based as an example on NaCl crystal has a value $\lambda_{1hf} = 0.6\mu m$ and lies in the optical range. Laser pulses in the infrared frequency range $(\lambda_L \approx 1\mu m)$ with pulse duration $\omega_L t_L \gg 1$ will not excite these waves coherently. However, using intense laser pulses ($a_0 = 10^2$-$10^3$, $t_L = 2fs$) of the wavelength in X-ray range (1-10keV), considered in [21, 22], it is possible to excite intense wake fields in ion dielectric media in the infrared and optical ranges, as for this case $1/\omega_{1hf} \gg t_L \gg 1/\omega_L$.

Wake field excitation on frequencies of the high-frequency branches of optical polarization electron oscillations $\omega_{Le\mp}$ by X-ray laser pulse [21] can be used for high-gradient acceleration similarly to laser-plasma wakefield acceleration [12-14,22].

## CONCLUSION

The process of excitation of wake Cerenkov radiation by a laser pulse in an ion dielectric waveguide is investigated. For definiteness, a diatomic ion crystal medium is considered. The nonlinear electric polarization of the ion dielectric medium, induced by the ponderomotive force of the laser pulse, is determined. The total electric polarization in the ion dielectric includes the electron polarization of the electron shells of ions of opposite charges, as well as the ion polarization caused by the displacement of ions in the electric field. A system of three strongly coupled linear oscillator equations is obtained, which describes the excitation of partial electric polarizations of an ion dielectric by a ponderomotive force of a laser pulse. The solution of these equations is obtained and the complete polarization in a diatomic ion dielectric medium is determined. Accordingly, expressions are obtained for polarization charges and currents, which, in turn, are the source of Cerenkov wake waves. The frequency spectrum and the space-time structure of the Cherenkov wake field, excited by a laser pulse in an ion dielectric waveguide, is determined. It is shown that in the infrared (microwave) frequency range, the excited wake electric field consists of a potential field of longitudinal optical phonons and a set of eigen wake electromagnetic waves of a dielectric waveguide. The dielectric constant in the infrared (microwave) frequency range in ion dielectrics always exceeds the dielectric constant in the optical range. Therefore, the condition of the Cherenkov radiation of a laser pulse in ion dielectrics is always satisfied.